# Electronic procrystalline state in moiré structures

Hui Guo[1,2,3#*], Zihao Huang[1,2#], Yixuan Gao[4#], Haowei Chen[5], Hao Zhang[1,2], Qian Fang[1,2], Yuhan Ye[1,2], Xianghe Han[1,2], Zhongyi Cao[1,2], Jiayi Wang[1,2], Runnong Zhou[1,2], Zhilin Li[1,2], Chengmin Shen[1,2,3], Haitao Yang[1,2,3], Hui Chen[1,2,3*], Wang Yao[5], Ziqiang Wang[6*], and Hong-Jun Gao[1,2,3*]

[1] Beijing National Center for Condensed Matter Physics and Institute of Physics, Chinese Academy of Sciences, Beijing 100190, PR China

[2] School of Physical Sciences, University of Chinese Academy of Sciences, Beijing 100190, PR China

[3] Hefei National Laboratory, Hefei 230088, China

[4] State Key Laboratory for Advanced Metals and Materials, University of Science and Technology Beijing, Beijing 100083, China

[5] New Cornerstone Science Laboratory, Department of Physics, The University of Hong Kong, Hong Kong, China

[6] Department of Physics, Boston College, Chestnut Hill, MA, USA

[#]These authors contributed equally to this work
[*]Correspondence to: hjgao@iphy.ac.cn, guohui@iphy.ac.cn, hchenn04@iphy.ac.cn, wangzi@bc.edu

**Solid state materials can display varieties of atomic structural orders ranging from crystalline to amorphous, underlying their properties and diverse functionalities. Procrystal has emerged as a new category of solids, featuring a long-range ordered lattice framework tiled with disordered atomic/molecular structures on the lattice sites, arousing great interest due to its novel structural and physical properties. However, the electronic analogue of a procrystal, dubbed as an electronic procrystalline (EPC) state, has never been experimentally observed. Here, we report the observation of an EPC state in a moiré superstructure formed between a monolayer metallic $NiTe_2$ and a superconductor $NbSe_2$ with incommensurate lattice wavevectors. The observed EPC state exhibits a long-range periodic charge modulation at the moiré scale inlaid with short-range irregular orders within each moiré cell. Strikingly, the short-range charge orders inside the moiré unit cells have proximately $\sqrt{3}\times\sqrt{3}$ quasi-period, which is absent in pristine $NiTe_2$. Intriguingly, the EPC order is also observed in the superconducting state of the moiré superstructure. Furthermore, the emergent EPC state and short-range charge order, coexisting with the proximity induced superconductivity, can be precisely modulated with the thickness of $NiTe_2$. Our findings uncover the potential of moiré platform for understanding and tuning novel correlated quantum phases with this exotic procrystalline order.**

The states of solid materials, broadly categorized by atomic arrangements into crystalline, amorphous, and the intermediate quasi-crystalline and liquid-crystalline, profoundly influence their physical properties and functional capabilities[1-5]. Analogously, electrons in crystals can also organize into their own distinct quantum states with the symmetry characteristics matching those of classical solid states, underlying exotic electronic properties. Of particular interest are the electronic liquid-crystalline (ELC) states in strongly-correlated systems, which break the point-group symmetry of the host crystal and serve as key platforms for studying correlated electron behaviors[6-8]. Notable examples include nematic states in high-temperature superconductors and heavy fermion systems, where the breaking of rotation symmetry results in anisotropic transport and offers insights into quantum criticality[9-13]. Similarly, smectic and stripe phases, characterized by the breaking of translation symmetry, are associated with charge density wave orders and unconventional pairing mechanisms[14-18]. Unveiling these electronic analogues of atomic structural orders fundamentally advances our understanding of a rich landscape of emergent quantum phases and exotic phenomena in condensed matter systems.

Recently, a kind of disordered crystal, procrystal has emerged as a new category of solid-state material, attracting significant interest due to its unique structural characteristics and potential physical properties inaccessible to conventional crystals[19-23]. The procrystal features a crystal-like aperiodic structure where disordered building blocks are arranged periodically in an ordered lattice potential but lacks crystalline translational symmetry at an atomic scale (Fig. 1**a**). This unique hybrid framework combines the attributes of both order and correlated disorder, presenting exciting opportunity towards emergent collective properties[23-27]. Analogously, the quantum electronic phase with characteristics of the procrystal, dubbed as the electronic procrystalline (EPC) state, can feature irregularly short-range-ordered electronic states as random building blocks to form long-range periodic lattice (Fig. 1**b**). Such unique electronic state EPC breaks the translational symmetries of the underlying crystal, precluding a Bloch description. Unveiling the EPC state promises profound insights into novel quantum phenomena in systems with such hybrid of long-range order and short-range disorders. However, this has yet to be experimentally realized. Here, we demonstrate the realization of an EPC state in a moiré superstructure between two metallic transition-metal dichalcogenides (TMDs) with incommensurate lattice wavevectors. The moiré superstructure is created by putting a monolayer metallic TMDs on a bulk metallic TMDs surface (Fig. 1**c**). This leads to the emergence of an EPC state characterized by large-periodic charge modulation at the moiré scale, accompanied by disordered or short-range-ordered electronic states inside the moiré unit cells (Fig. 1**d**).

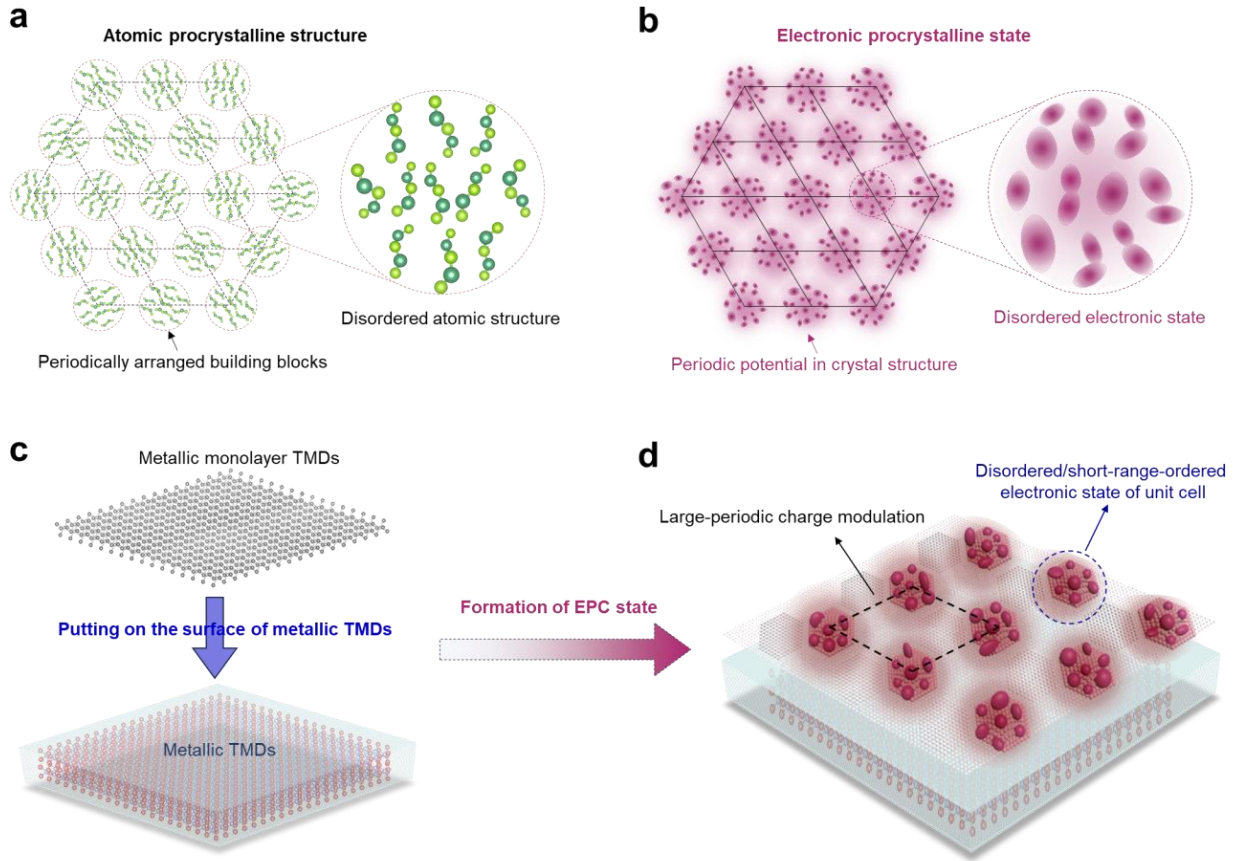

**Fig. 1. Schematic of atomic procrystalline structure and electronic procrystalline (EPC) state. a**, Illustration of a procrystalline structure, depicting a crystal-like aperiodic structure where disordered building blocks are arranged periodically in an ordered lattice but lacks the translational symmetry at an atomic scale. **b,** EPC state, representing a disordered electronic state in the presence of periodic potential in a crystal structure. **c,** Formation of a moiré superstructure by putting a monolayer metallic TMDs on a bulk metallic TMDs surface. **d,** Formation of an EPC state with large-periodic charge modulation and disordered/short-range-ordered electronic states of moiré unit cells.

We start with constructing the moiré structure by direct epitaxial growth of metallic monolayer (1L) $NiTe_2$ film on a superconducting $NbSe_2$ substrate (Fig. 2**a**). Sharp diffraction spots of the low-energy electron diffraction (LEED) pattern indicates the large-scale ordered moiré superstructure of the $NiTe_2/NbSe_2$ (Fig. 2**b**). Two sets of the LEED pattern of the $NiTe_2$ (red circles) and $NbSe_2$ (blue circles) indicate that monolayer NiTe$_2$ share the same orientation with NbSe$_2$. The different length of the two LEED patterns is due to lattice-mismatch, leading to the formation of the moiré superstructure. Large-periodic charge modulation at the moiré scale is observed with high-resolution STM (Fig. 2c), which features three inequivalent local regions within a moiré unit cell, i.e., the bright (AB$_{Hollow}$), dark (AA), and intermediate (AB$_{Se}$) regions, respectively (see Methods). Remarkably, electronic structures at these three local regions exhibit strong intensity contrast, with the brightest AB$_{Hollow}$ regions appearing as bubbles, suggesting a significant charge density modulation. The d$I$/d$V$ spectra of the three regions in NiTe$_2$ further corroborate (Fig. 2d,e) the charge density modulation. The d$I$/d$V$ maps (Fig. 2f and S1) reveal distinct inversion of intensity contrast at +250 meV and -250 meV at the same spatial location, further confirming the large-periodic charge density modulation at the moiré scale[28,29].

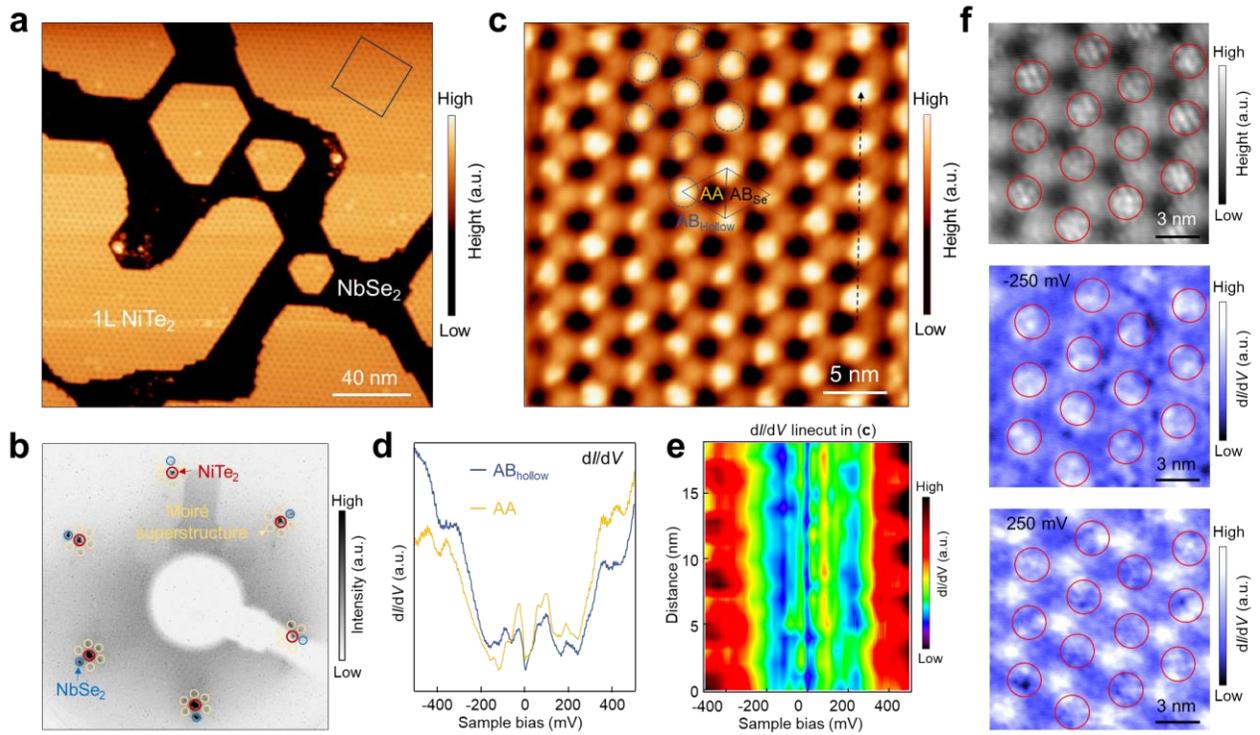

**Fig. 2. Demonstration of large-periodic charge modulation in NiTe$_2$/NbSe$_2$. a,** A large-scale STM image ($V_s$=-2.0 V, $I_t$=0.05 nA), showing a moiré pattern of the 1L NiTe$_2$ on the NbSe$_2$. **b,** LEED pattern, showing the formation of large-scale moiré superstructure (yellow circles) with sharp diffraction spots from NiTe$_2$ (red circles) and NbSe$_2$ (blue circles). **c,** A zoom-in STM image ($V_s$ = -990 mV, $I_t$ = 0.03 nA) taken at the black-box region in (a), showing a bubble-like large-periodic charge modulation at the moiré scale. The black rhombus marks a unit cell of the superstructure, highlighting the three inequivalent local regions AB$_{Hollow}$, AA and AB$_{Se}$ inside the unit cells. **d,** The spatially-averaged d$I$/d$V$ spectra

at the bright AB$_{Hollow}$ (blue) and dark AA (green) regions ($V_s$ = -0.5 V, $I_t$ = 1.0 nA, $V_{mod}$ = 5 mV), showing a significant variation of the electronic states in the two regions. **e**, The d$I$/d$V$ line-cut across the bubble regions (along the arrow in (**c**)), showing the modulation of the electronic states at the moiré scale ($V_s$ = -0.5 V, $I_t$ = 1.0 nA, $V_{mod}$ = 5 mV). **f**, The topography (top) and corresponding d$I$/d$V$ maps at -250 meV (middle) and 250 meV (bottom), respectively, showing the inversion of intensity contrast for the bubble regions (red circles) ($I_t$ = 1.0 nA, $V_{mod}$ = 5 mV).

Next, we demonstrate the short-range charge order inside each moiré unit cell. The short-range-ordered electronic states are observed at each AB$_{Hollow}$ regions of the moiré superstructure (Fig. 3**a** and S2). The sharp spots of the fast Fourier transform (FFT) pattern (Fig. 3**b**) indicates the Bragg lattice of NiTe$_2$ and the ordered superstructure of NiTe$_2$/NbSe$_2$. The broad spots surrounding six hexagonal wavevectors at ($\sqrt{3}\times\sqrt{3}$)$R$30° positions ($Q_{\sqrt{3}\times\sqrt{3}}$) with respect to NiTe$_2$ Bragg lattice suggest the formation of short-range order at approximately $\sqrt{3}\times\sqrt{3}$ of the NiTe$_2$ unit cell. The short-range orders are irregular, varying between different AB$_{Hollow}$ regions, as evident by the distribution statistics of rotation angle and periodicity (Fig. 3**c,d**). Thus, the overall charge density modulation has the precise character of procrystalline order, with long-range periodic order pinned by the moiré landscape, whereas the short-range orders within each moiré cell exhibit quasi-hexatic characteristics, with quasi-sixfold rotational symmetry and a quasi-periodicity of approximately $\sqrt{3}$ times of the NiTe$_2$ lattice constant.

We further analyze the short-range charge order state inside the moiré unit cells. At the same spatial locations, the density of states (DOS) shows contrast inversion on opposite sides of $E_F$ (+100 meV and -100 meV) [28,29] (Fig. 3**e,f** and S3), with the same short-range $\sqrt{3}\times\sqrt{3}$ orders. The spatial distribution of the amplitude and phase of the $Q_{\sqrt{3}\times\sqrt{3}}$ reveals the short-range charge order within the AB$_{Hollow}$ regions with phase variation (Fig. 3**g,h**). The formation of the short-range $\sqrt{3}\times\sqrt{3}$ charge order likely originate from the enhanced electron correlation driven by the cooperative effect of moiré-confined strain, charge transfer and localization of electron density at the AB$_{Hollow}$ regions (Fig. S4-S9, see Methods). These results demonstrate the emergence of short-range $\sqrt{3}\times\sqrt{3}$ charge orders inside the moiré unit cells, while it breaks the underlying crystal symmetries and losses long-range phase coherence.

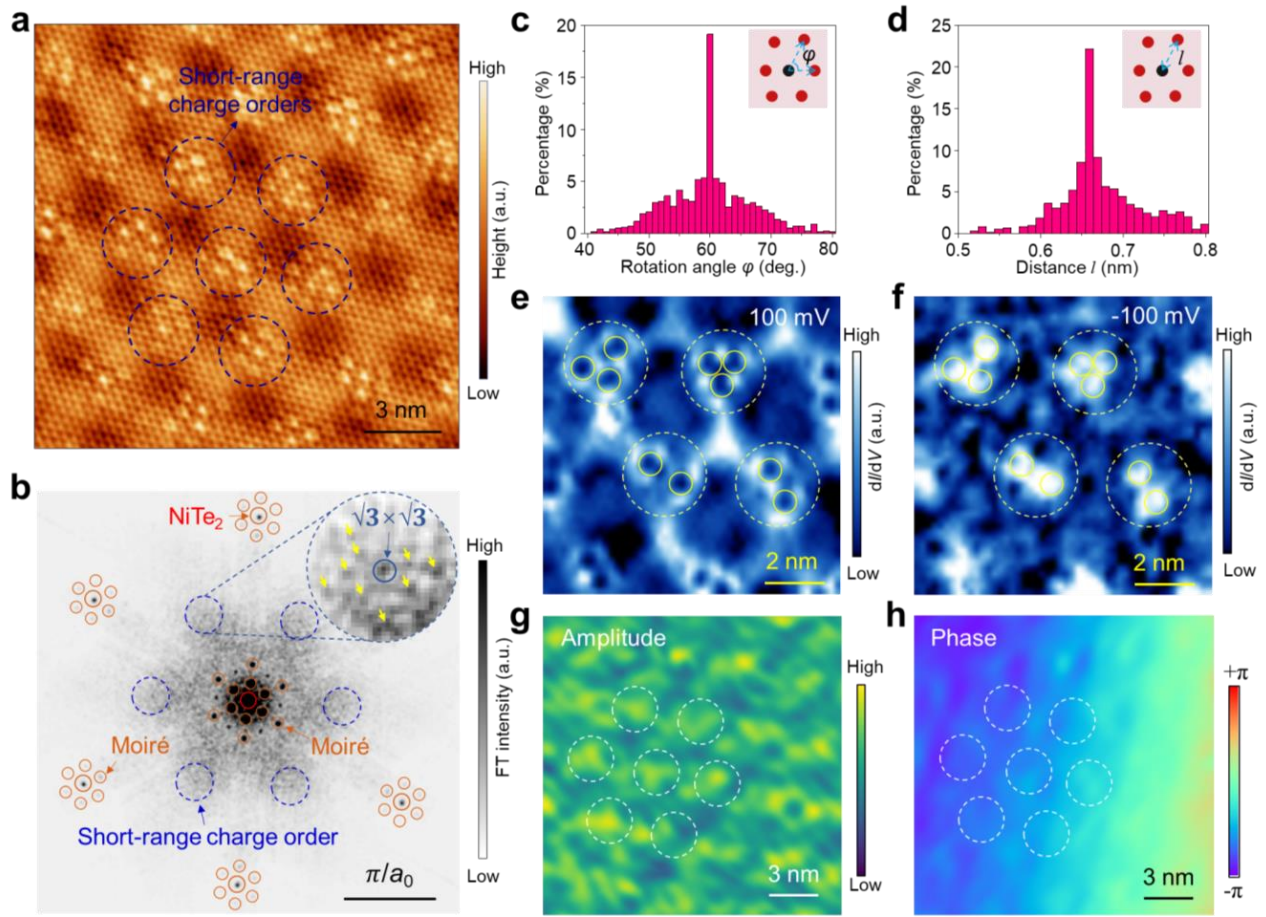

**Fig. 3. Demonstration of short-range charge orders inside the moiré unit cells. a,** An atomic-resolution STM image, showing the moiré superstructure, the atomic lattice of NiTe$_2$, and short-range ordered states within the AB$_{Hollow}$ regions ($V_s$ = -50 mV, $I_t$ = 5 nA). **b,** The symmetrized Fourier transform pattern, showing sharp diffraction spots from the NiTe$_2$ Bragg lattice (red circles), the ordered moiré superstructure (orange circles), and a set of broad spots (yellow arrows) around ($\sqrt{3}\times\sqrt{3}$)$R30°$ positions (blue arrow) associated with the short-range ordered states (blue circles). **c,d,** Distribution statistics of the rotation angle $\varphi$ (**c**) and distance $l$ (**d**) of the short-range ordered states, revealing quasi-sixfold rotational symmetry and a quasi-periodicity of approximately $\sqrt{3}$ times the NiTe$_2$ lattice constant. Insets define the rotation angle $\varphi$ and distance $l$. **e,f,** The d$I$/d$V$ maps ($I_t$ = 1.0 nA, $V_{mod}$ =5 mV) at +100 meV (**e**) and -100 meV (**f**), respectively, showing an inversion in intensity contrast. The AB$_{Hollow}$ regions and the internal short-range charge order are highlighted by the red and yellow circles, respectively. **g,h,** Spatial distribution of the amplitude (**g**) and phase (**h**) corresponding to the ($\sqrt{3}\times\sqrt{3}$)$R30°$ vector with inverse-FFT of (**b**), showing that the charge order mainly occur within the AB$_{Hollow}$ regions (white circles) with variation of the phases.

Intriguingly, the EPC order is also observed in the superconducting state. The superconducting phase in NiTe$_2$ is induced by proximity effect (Fig. 4**a,b**). The coherence peak height, peak-to-peak distance, and consequently the superconducting gap size show strong modulation across the AB$_{Hollow}$ regions (Fig. 4**c**).

Within these regions, the coherence peak height is enhanced, while the superconducting gap size is reduced. This spatially-modulated pair density represents a secondary phase of the charge density modulation at the moiré scale, as a result of the correlation between the EPC state and the proximity-induced superconductivity. The periodicity of the pair density modulation is obviously variable and the d$I$/d$V$ map at the energy of coherence peak shows inhomogeneous characteristics (Fig. 4**d,e**), suggesting a formation of paired electronic procrystalline state and reinforcing the disordered nature of the electronic state in such moiré system.

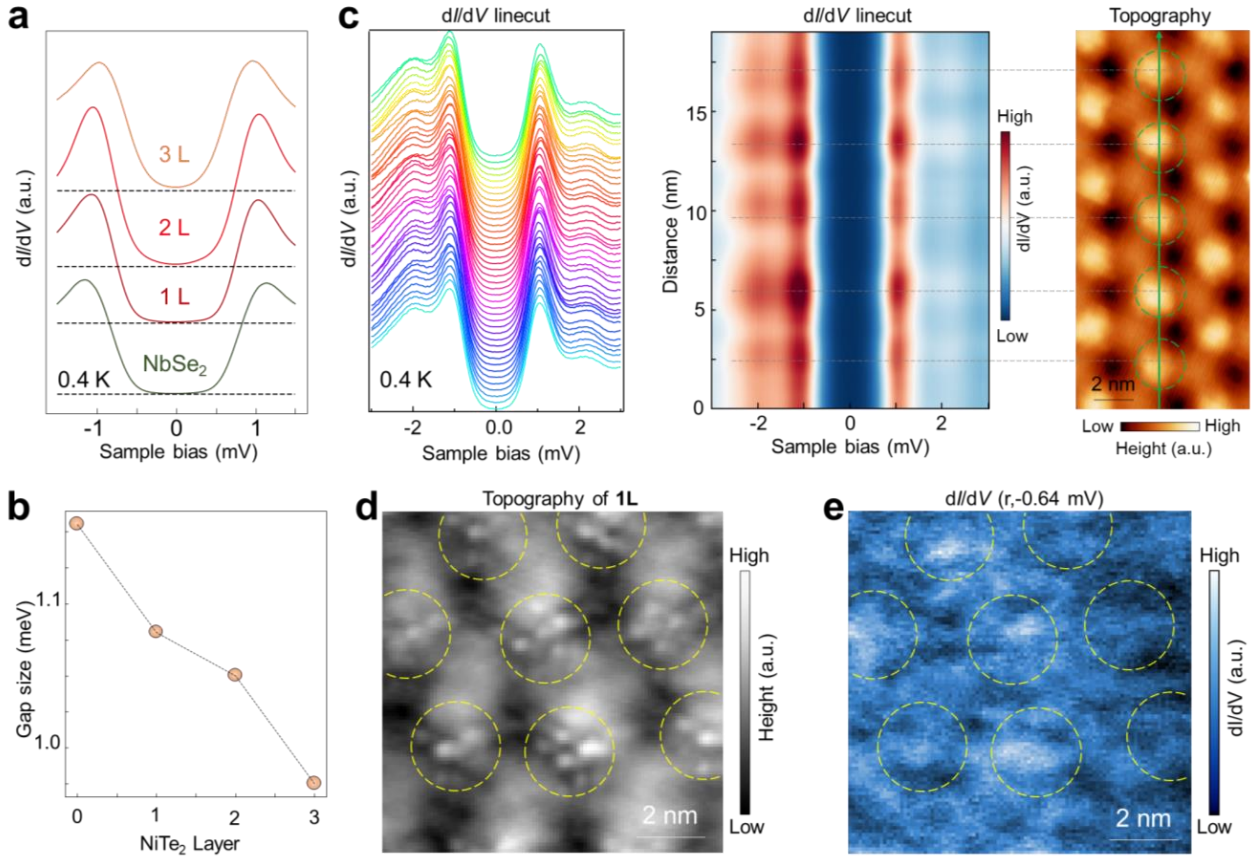

**Fig. 4. Demonstration of the EPC order in the superconducting state of NiTe$_2$. a**, The spatially-averaged d$I$/d$V$ spectra ($V_s$ = -3 mV, $I_t$ = 1 nA, $V_{mod}$ = 0.05 mV) at NbSe$_2$ substrate, monolayer (1L), bilayer (2L), and trilayer (3L) NiTe$_2$, respectively, showing broadening two coherence peaks and increasing in-gap conductance with the layer of NiTe$_2$. **b**, The SC gap size subtracted from Dyen function fitting from d$I$/d$V$ spectra as a function of the layer of NiTe$_2$ (bare NbSe$_2$, L=0), showing a monotonic decrease with increasing layers, indicating the proximity effect. **c**, The waterfall and intensity plot of the d$I$/d$V$ linecut ($V_s$ = -3 mV, $I_t$ = 1 nA, $V_{mod}$ = 0.05 mV) across AB$_{Hollow}$ regions (dash circles) along the green arrow in the topography, showing strong modulation of superconducting coherence peak and gap size, indicating a pair

density modulation. **d,e,** STM topography of the monolayer NiTe$_2$ and corresponding d$I$/d$V$ map at -0.56 meV, showing inhomogeneous superconductivity and the enhancement of in-gap states in AB$_{Hollow}$ regions (yellow dotted circles) ($V_s$ = -2 mV, $I_t$ = 1 nA, $V_{mod}$ = 0.05 mV), suggesting a formation of paired electronic procrystalline state.

Finally, we show the modulation of the short-range charge order and EPC state by tuning the thickness of NiTe$_2$ (Fig. S10). In comparison to the 1L NiTe$_2$ EPC state, in the bilayer (2L), the moiré modulation remains while the √3×√3 short-range charge order is suppressed (Fig. **5a**). The density of states in equivalent local regions, e.g., AB$_{Hollow}$, exhibit evident inhomogeneity, indicating that the electronic state still retains disordered nature. The d$I$/d$V$ maps further reveal the distinct disordered electronic states within different moiré regions and weaker charge modulation than that of the 1L NiTe$_2$ at the moiré scale (Fig. **5b,c** and S11), implying a weak EPC state without √3×√3 charge orders in the bilayer NiTe$_2$. As the thickness increases to trilayer, the charge modulation are further suppressed and eventually disappeared at the four-layer NiTe$_2$ (Fig. **5d,e**). Figure **5f** shows a lateral heterostructure of 2L NiTe$_2$/NbSe$_2$ and 1L NiTe$_2$/NbSe$_2$. The corresponding d$I$/d$V$ maps reveal large-periodic charge modulation with inlaid short-range charge order in the 1L NiTe$_2$ region, whereas the 2L exhibits much weaker modulation without charge orders (Fig. **5g** and S12), indicating the formation of a junction between the EPC states with and without √3×√3 short-range charge orders. A pronounced charge edge state at about 45 meV emerges at the interface (Fig. **5h**), which is further confirmed by a d$I$/d$V$ linecut across the 1L-2L junction (Fig. **5i**). Notably, such edge state emerges at the interface of 1L-2L junction, but is absent at the 2L-3L junction (Fig. S13).

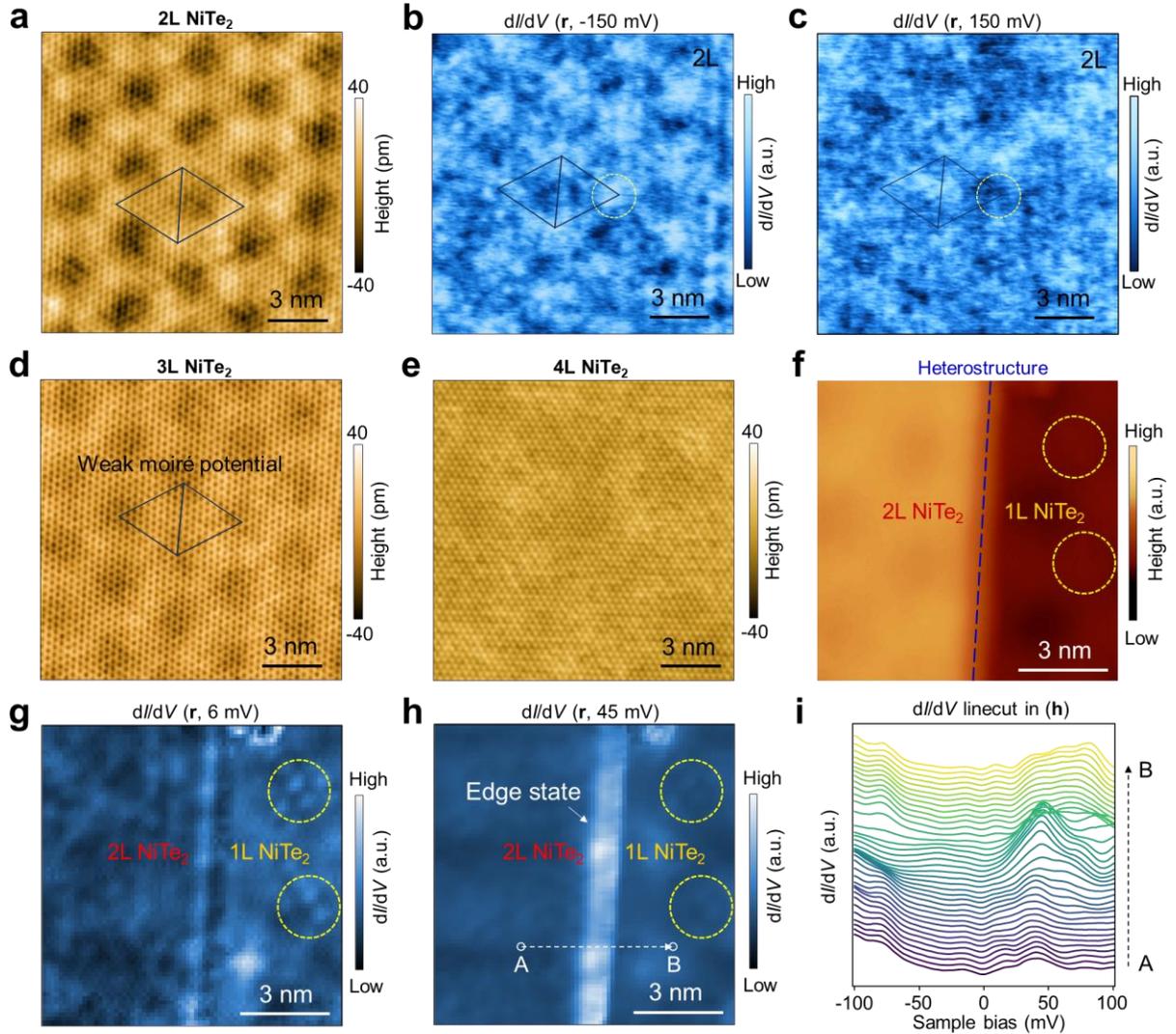

**Fig. 5. Modulation of the short-range charge order and EPC state. a,** Bilayer (2L) NiTe$_2$ on NbSe$_2$ (STM image, $V_s$ = -10 mV, $I_t$ = 5 nA), showing large-periodic charge modulation at the moiré scale without short-range charge order. **b,c,** The d$I$/d$V$ maps of the 2L NiTe$_2$ at -150 meV (**b**) and +150 meV (**c**) ($I_t$ = 1.0 nA, $V_{mod}$ = 5 mV), showing distinct disordered electronic states within different moiré regions and weaker modulation than that of the monolayer (1L) NiTe$_2$ at the moiré scale. **d,e,** Atomically-resolved STM images of the trilayer (3L) and four-layer (4L) NiTe$_2$ ($V_s$ = -10 mV, $I_t$ = 5 nA), respectively, showing that the moiré potential and EPC state are gradually suppressed with increasing layers. **f,** Heterostructure formation between 1L and 2L NiTe$_2$ (STM image, $V_s$ = -100 mV, $I_t$ = 80 pA), showing the boundary between the 1L region with short-range √3×√3 charge order (yellow circles) and the 2L region without short-range charge order. **g,h,** The corresponding d$I$/d$V$ maps of (**f**) ($I_t$ = 1.0 nA, $V_{mod}$ = 1 mV) at 6 meV (**g**) and 45 meV (**h**), respectively, showing the difference of the EPC state in 1L and 2L regions and the formation of a charge edge state at the interface. **i,** The waterfall plot of d$I$/d$V$ linecut along the white dash arrow in (**h**), showing the edge state of the heterostructure at about 45 mV.

In summary, we have discovered a quantum-textured EPC state in a moiré structure between metallic NiTe$_2$ and superconducting NbSe$_2$. The EPC state shows a long-range periodic charge modulation in the monolayer NiTe$_2$ at the moiré scale, with inlaid disordered electronic state inside the moiré unit cells exhibiting short-range √3×√3 order. We show that the EPC can coexist with the proximity induced superconductivity NiTe$_2$, and the short-range order can be modulated with the thickness of NiTe$_2$. This further points to exiting possibilities and fine tunability to explore the interplay of the EPC charge order with superconductivity, and potentially other correlation phenomena such as spin density waves[30], pair density waves[31], intertwined density waves orders[32]. Our findings open the door towards a new class of correlated electronic states that can generally exist in moiré systems formed by metallic layers with pronounced carrier density, not available in the extensively studied twisted semimetallic graphene and semiconducting TMDs systems[33-53].

**Methods**

**Sample preparation.** NiTe₂ film with various thickness were epitaxially grown on freshly-cleaved NbSe₂ substrates by molecular beam epitaxy approach under ultra-high vacuum conditions (UHV, base pressure ~5 × 10⁻¹⁰ mbar). High-quality NbSe₂ single crystals used as substrates were fabricated by chemical vapor transport approach. The clean NbSe₂ substrates were obtained by cleaving in vacuum and subsequently annealing in UHV at 600 K for several hours. The NiTe₂ films were grown by e-beam evaporation of Ni (99.9%, *Goodfellow Cambridge Ltd.*) and simultaneous deposition of atomic Te (99.99%, *Sigma-Aldrich*) from a Knudsen cell at a substrate temperature of 520 K. During growth, the Te flux approximately an order of magnitude greater Ni flux (Se-rich conditions) is used.

**STM/STS and LEED.** STM/STS measurements were performed in an ultrahigh vacuum (1×10⁻¹⁰ mbar) ultra-low temperature condition equipped with 11 T magnetic field. The stable sample temperature can be kept at a base temperature of 0.4 K and 4.2 K, respectively. The electronic temperature is 620 mK at a base temperature of 400 mK. All the scanning parameters (setpoint voltage and current) of the STM topographic images are listed in the captions of the figures. Unless otherwise noted, the differential conductance (d$I$/d$V$) spectra were acquired by a standard lock-in amplifier at a modulation frequency of 973.1 Hz. Tungsten tip was fabricated via electrochemical etching and calibrated on a clean Au(111) surface prepared by repeated cycles of sputtering with argon ions and annealing at 770 K. LEED was employed with a 4-grid detector (*Omicron Spectra LEED*) in the UHV chamber at room temperature.

**Two-dimensional lock-in technique.** To explore the √3×√3 charge modulations, we employ a two-dimensional lock-in technique[54] to determine the amplitude and phase of the modulations. For any arbitrary real space image:

$$A(\mathbf{r}) = \sum_{\mathbf{Q}} a_{\mathbf{Q}}(\mathbf{r}) e^{-i\mathbf{Q}\cdot\mathbf{r}}$$

where $a_{\mathbf{Q}}(\mathbf{r})$ is the complex amplitude at wavevector Q and position r. If Q is the wavevector of interest, it can be extracted from the Fourier transform $A(\mathbf{q})$ by shifting it back to the center and multiplying a Gaussian window with a cut-off length σ in q-space. The approximate complex amplitude in real space $A_{\mathbf{Q}}(\mathbf{r})$ can be obtained by inverse Fourier transform as following:

$$A_{\mathbf{Q}}(\mathbf{r}) = F^{-1}[A_{\mathbf{Q}}(\mathbf{q})] = \int d\mathbf{R} A(\mathbf{R}) e^{i\mathbf{Q}\cdot\mathbf{R}} e^{-\frac{(r-R)^2}{2\sigma^3}}$$

Thus, using this technique, the amplitude $|A_{\mathbf{Q}}(\mathbf{r})|$ and spatial phase $\Phi_{\mathbf{Q}}^A(\mathbf{r})$ of the modulation at Q can be written as:

$$|A_{\mathbf{Q}}(\mathbf{r})| = \sqrt{\text{Re } A_{\mathbf{Q}}(\mathbf{r})^2 + \text{Im } A_{\mathbf{Q}}(\mathbf{r})^2}$$

$$\Phi_{\mathbf{Q}}^A(\mathbf{r}) = \tan^{-1} \frac{\text{Im } A_{\mathbf{Q}}(\mathbf{r})}{\text{Re } A_{\mathbf{Q}}(\mathbf{r})}$$

**Definition of $AB_{Hollow}$, $AB_{Se}$, and AA regions in the moiré structure of $NiTe_2/NbSe_2$.**

To determine the local atomic structure of the bright, dark, and intermediate regions within the moiré superstructure, we construct a structural model of monolayer $NiTe_2$ on $NbSe_2$. After structural optimization, the moiré superstructure shows three distinct local regions, denoted as $AB_{Hollow}$, $AB_{Se}$, and AA based on different atomic alignments (Fig. S14). In the $AB_{Hollow}$ region, Ni atoms are positioned on top of hollow sites in $NbSe_2$ substrate, while in the $AB_{Se}$ and AA regions, Ni atoms are aligned above Se and Nb atoms, respectively. the DFT simulated STM image based on the optimized atomic model shows excellent agreement with the experimental data (Fig. S15). Consequently, we attribute the bright, dark, and intermediate regions within the moiré unit cells to $AB_{Hollow}$, AA, and $AB_{Se}$ regions, respectively.

**Possible origin of the short-range charge order within the $AB_{Hollow}$ regions.**

To study the possible origin of the short-range charge order within $AB_{Hollow}$ regions of the monolayer $NiTe_2$, we analyzed the moiré-confined strain distribution[55], charge transfer, and electron density distribution in the $NiTe_2$ on $NbSe_2$ with DFT calculations. The charge order is absent in the pristine $NiTe_2$[56,57], and the phonon dispersion of monolayer $NiTe_2$ shows no phonon softening, indicating its thermodynamic stability (Fig. S4). However, in the moiré structure of $NiTe_2/NbSe_2$, the monolayer $NiTe_2$ within $AB_{Hollow}$ regions experiences significant compressive strain and shows a strong localization of the electronic states (Fig. S5). The band structure upon applying a 10% compressive strain (Fig. S6) shows that the energy band around the K point anchored by the Dirac point is pulled closer to the Fermi level, leading to the formation of a small hole pocket located at K point precisely, and thereby enhancing its density of states (Fig. S7). Moreover, the maximum electronic susceptibility is moved to K point, which indicates the nesting between K and K′ points, suggesting an enhanced electron correlation. As a result, the Fermi surface nesting vector gives rise to a $\sqrt{3}\times\sqrt{3}$ charge order vector. The atomic structure model for the $\sqrt{3}\times\sqrt{3}$ phase shows a lower symmetry ($C_{3v}$) compared to the high-symmetry of the pristine phase ($D_{3d}$) (Fig. S8). Furthermore, both the strain and the localization of electron density within $AB_{Hollow}$ regions of the bilayer $NiTe_2$ show much smaller than that in the monolayer $NiTe_2$ (Fig. S9), which fails to induce a charge order in the bilayer $NiTe_2$. Therefore, we speculate that the formation of the short-range charge order originate mainly from the enhanced electron correlations driven by the cooperative effect of moiré-confined strain, charge transfer, and localization of electron density in the $AB_{Hollow}$ regions.

**Discussion of the significance of the disordered electronic state EPC.**
The EPC, disordered electronic state, in the metallic moiré structure of $NiTe_2/NbSe_2$ goes well beyond prior studies on twisted semimetallic graphene and semiconducting TMDs[33-50], where the low-energy electronic states can be described by translation invariant Hamiltonian[51,53]. In contrast, in the moiré structure between two metallic TMDs with incommensurate lattice wavevectors, the conventional periodic Hamiltonian models become unfeasible[58]. The electrons experience competing and incommensurate potentials, leading to the loss of translational symmetry and breakdown of periodic band structures[59], which is likely a key factor enabling the emergence of the EPC state in the moiré structure of $NiTe_2/NbSe_2$. This discovery is expected to inspire new insights into understanding novel topological and correlated quantum phenomena in moiré superlattices beyond translation invariant Hamiltonian description.

**DFT calculations.** The DFT calculations were carried out using Vienna ab initio simulation package (VASP)[60]. The projected augmented wave (PAW) method[61] was used to describe the core-valence interactions. The generalized gradient approximation (GGA)[62] in the form of Perdew-Burke-Ernzerhof (PBE) is adopted for the exchange-correlation functional. In the calculations of free-standing $NiTe_2$, a 15 Å vacuum layer is used, and all atoms are fully relaxed until the residual forces on each atom are smaller than 0.01 eV/Å. The k-points sampling is $27 \times 27 \times 1$ with the Gamma scheme. Wave functions were expanded on a plane-wave basis set up to 500 eV energy cutoff. A slab model is used for the moiré superlattice. Wave functions were expanded on a plane-wave basis set up to 400 eV energy cutoff. Van-der-Waals interactions were considered at the DFT-D3 level. The vacuum layer is larger than 10 Å. A Γ-only k-point sampling in the first Brillouin zone is used for the moiré superlattice. All atoms except the NbSe$_2$ substrate are fully relaxed until the net force is smaller than 0.01 eV/Å. The in-plane strain is introduced by changing the lattice constant of NiTe$_2$. The strain is defined by ΔL/L, where L and ΔL are the lattice constant and compressed length of pristine NiTe$_2$. The real part of susceptibility is calculated as the following formula,

$$x'_0(q) = \sum_k \frac{f(\epsilon_{k+q}) - f(\epsilon_k)}{\epsilon_k - \epsilon_{k+q}}$$

where the *f* denotes the electron's Fermi distribution function.

## Data availability
Data measured or analyzed during this study are available from the corresponding author on reasonable request.

## Acknowledgements
We thank Min Ouyang for valuable discussions. This work is supported by grants from the National Natural Science Foundation of China (62488201), the National Key Research and Development Projects of China (2022YFA1204100), the CAS Project for Young Scientists in Basic Research (YSBR-053 and YSBR-003) and the Innovation Program of Quantum Science and Technology (2021ZD0302700). Z.W. is supported by U.S. Department of Energy, Basic Energy Sciences Grant DE-FG02-99ER45747.

**Author Contributions:** H.-J.G. supervised the project. H.-J.G., H.C., and H.G. designed the experiments. H.G., R.Z., J.Y.W., and Q.F. fabricated the samples. Z.L., H.T.Y., and C.M.S. synthesized the high-quality NbSe$_2$ bulk crystals. H.G., Z.H., H.C., Y.Y., X.H., H.Z., and Z.C. performed STM experiments. Y.G.,